\documentclass{article}
\pdfoutput=1 
\usepackage[english]{babel}
\usepackage[utf8]{inputenc}
\usepackage[colorlinks]{hyperref}
\usepackage{enumitem,kantlipsum}
\usepackage{graphicx, color}
\usepackage{wrapfig}
\usepackage{amsfonts}

\newtheorem{proposition}{Proposition}

\title{{\Huge PACT\/}:   {\Huge P\/}rivacy-Sensitive Protocols {\Huge A\/}nd Mechanisms
\\for Mobile {\Huge C\/}ontact {\Huge T\/}racing }
\date{}

\author{
Justin Chan$^{1}$, Dean Foster$^{5}$, Shyam Gollakota$^{1}$, Eric Horvitz$^{\dag,1,3,4}$,  Joseph Jaeger$^{\dag,1}$,\\ 
Sham Kakade$^{\dag,1,2}$, Tadayoshi Kohno$^{1}$, John Langford$^{\dag,4}$, Jonathan Larson$^{4}$, Puneet Sharma$^{6}$,\\   Sudheesh Singanamalla$^{1}$,
Jacob Sunshine$^{3}$, Stefano Tessaro$^{\dag,1}$  
\\
\\
	$^1$ Paul G. Allen School of Computer Science \& Engineering, University of Washington
\\
	$^2$ Department of Statistics, University of Washington
\\
	$^3$ School of Medicine, University of Washington
\\
	$^4$ Microsoft Research
\\
	$^5$ University of Pennsylvania
\\
    $^6$ Boston Public Health Commission
\\
	$^{\dag}$ Corresponding authors\footnote{Email: \texttt{jsjaeger@cs.washington.edu}, \texttt{eric@horvitz.org}, \texttt{sham@cs.washington.edu}, \texttt{jl@hunch.net},
	\texttt{tessaro@cs.washington.edu}
	}
}

\begin{document}

\maketitle

\begin{abstract}
The global health threat from COVID-19 has been controlled in a number of instances by large-scale testing and contact tracing efforts.
We created this document to suggest three functionalities on how we might best harness computing technologies to supporting the goals of public health organizations in minimizing morbidity and mortality associated with the spread of COVID-19, 
while protecting the civil liberties of individuals.
In particular, this work advocates for a third-party--free approach to assisted mobile contact tracing, because such an approach mitigates the security and privacy risks of requiring a trusted third party. We also explicitly consider the inferential risks involved in any contract tracing system, where any alert to a user could itself give rise to de-anonymizing information.  

More generally, we hope to participate in bringing together colleagues in industry, academia, and civil society to discuss and converge on ideas around a critical issue rising with attempts to mitigate the COVID-19 pandemic.
\end{abstract}

\section{Introduction and Motivation}

Several communities and nations seeking to minimize death tolls from COVID-19, are resorting to \emph{mobile-based, contact tracing technologies} as a key tool in mitigating the pandemic. Harnessing mobile computing technologies is an obvious means to dramatically scale-up conventional epidemic response strategies to do tracking at population scale. However, straightforward and well-intentioned contact-tracing applications can invade personal privacy and provide governments with justification for data collection and mass surveillance that are inconsistent with the civil liberties that citizens will and should expect---and demand. To be effective, acceptable, and consistent with the need to observe commitments to privacy, we must leverage designs and computing advances in privacy and security. In cases where it is valuable for individuals to share data with others, systems must provide voluntary mechanisms in accordance with ethical principles of personal decision making, including disclosure, and consent. We refer to efforts to identify, study, and field such privacy-sensitive technologies, architectures, and protocols in support of mobile tracing as PACT (\emph{P}rivacy sensitive protocols \emph{A}nd mechanisms for mobile \emph{C}ontact \emph{T}racing).

\begin{center}
\emph{The objective of PACT is to set forth transparent privacy and
  anonymity standards,\\
  which permit adoption of mobile contract tracing efforts while upholding civil liberties.}
\end{center}

\begin{wrapfigure}{R}{0.4\textwidth}
\centering
\includegraphics[width=0.4\textwidth]{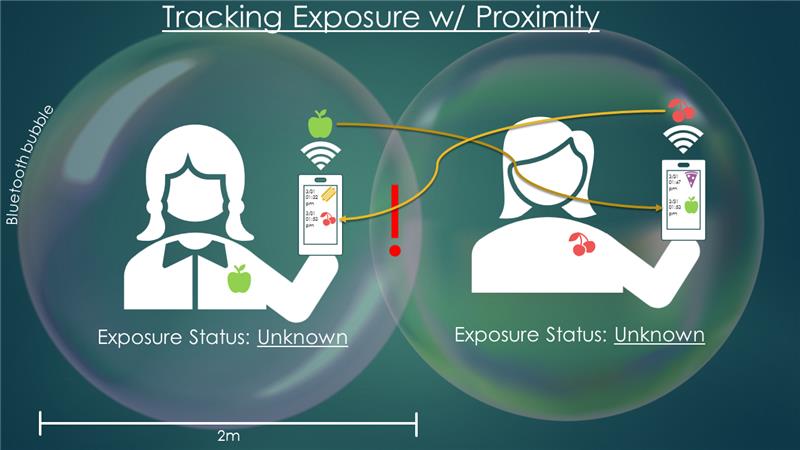}
\caption{\label{fig:bubble} Proximity-Based Tracing. The basic idea is that users broadcast signals (``pseudonyms''), while also recording the signals they receive. Notably, this \emph{co-location} approach
avoids the need to collect and share \emph{absolute} location information.
Credit: M Eifler.}
\end{wrapfigure}

This work specifies a third-party--free set of protocols and mechanisms in order to achieve these objectives. While approaches which rely on trusted third parties can be straightforward, many naturally oppose the aggregation of information and power that it represents, the potential for misuse by a central authority, and the precedent that such an approach would set.

It is first helpful to review the conventional contact tracing strategies executed by public health organizations, which operate as follows: Positively tested citizens are asked to reveal (voluntarily, or enforced via public health policy or by law depending on region) their contact history to public health officers. The public health officers then inform other citizens who have been at risk to the infectious agent based on co-location, via some definition of co-location, supported by look-up or inference about locations. The citizens deemed to be at risk are then asked to take appropriate action (often to either seek tests or to quarantine
themselves and to be vigilant about symptoms).  It is important to emphasize that the current approach \emph{already} makes a tradeoff between the privacy of a positively tested individual and the benefits to society.

We describe mobile contact-tracing functionalities that seeks to augment the services provided by public health officers, by enabling the following capabilities via computing and communications technology:

\begin{itemize}
\item \textbf{Mobile-assisted contact tracing interviews:}  A citizen who becomes ill can use this functionality to improve the efficiency and completeness of manual contact tracing interviews.  In many situations, the citizen can speed up the interview process by filling in much of a contact interview form before the contact interview process even starts, reducing the burden on public health authorities.  The privacy-sensitivity here is ensured since all the data remains on the user's device, except for what they voluntarily decide to reveal to health authorities in order to enable contact tracing. In advance of their making a decision to share, they are informed about how their data may be used and the potential risks of sharing. 

\item \textbf{Narrowcast messages:}  Public health authorities can make available custom-tailored messages to specific, relevant subsets of citizens.  For example, the following message might be issued: ``If you visited the X Eldercare Center between March 7th and 10th, please email yy@hhhealth.org'' or ``Please refrain from entering playground Z until April 6th because it needs to undergo decontamination.''  A mobile app can download all of these messages and display those relevant to a citizen based on the app's sensory log or potential future movements.  This capability allows public health officials to quickly warn people when new hotspots arise, or canvas for general information.  It enables a citizen to be well-informed about extremely local pandemic-relevant events.  

\item \textbf{Privacy-sensitive, mobile tracing:} Proximity-based signals seem to provide the best available contact sensor from one phone to another; see Figure~\ref{fig:bubble} for the basic approach. Proximity-based sensing can be done in a privacy-sensitive manner. With the approach, no absolute location information is collected nor shared. Variants of proximity-based analyses have been employed in the past for privacy-sensitive analyses in healthcare \cite{white2012}. Taking advantage of proximity-based signals can speed the process of contact discovery and enable contact tracing of otherwise undiscoverable people like the fellow commuter on the train.  This can also be done with a third-party--free approach providing similar privacy tradeoffs as manual contact tracing.  
This functionality can enable someone who has become ill with symptoms consistent with COVID-19, or who has received confirmation of infection with a positive test for COVID-19, to voluntarily and under a pseudonym, share information that may be relevant to the wellness of others.  In particular, a system can manage, in a privacy-sensitive manner, data about individuals who came in close proximity to them over a period of time (e.g., the last two weeks), even if there is no personal connection between these individuals.
 Individuals who share information do so with disclosure and consent around potential risks of private information being shared.  We further discuss disclosure, security concerns, and re-identification risks in Section~\ref{sect:FAQ}.
\end{itemize}

Importantly, these protocols, by default, keep \emph{all} personal data on a citizens' phones (aside for pseudonymous identifiers broadcast to other local devices), while enabling these key capabilities; information is shared via voluntary disclosure actions taken, with the understandings relayed via careful disclosure. For example, if someone never tests positive for COVID-19 or tests positive but decides not to use the system, then *NO* data is ever sent from their phone to any remote servers; such individuals
would be contacted by standard contact tracing mechanisms arising from reportable disease rules. The data on the phone can be encrypted and can be set up to automatically time out based on end-user controlled policies.  This would prevent the dataset from being accessed or requested via legal subpoena or other governmental programs and policies. 

 We specify protocols for all three separate functionalities above, and each app designer can decide which ones to use.  These protocols notably have different value adoption curves: Narrowcast and Mobile-assisted contact tracing have a value which is linear in the average adoption rate while privacy-sensitive mobile tracing has value quadratic in the average adoption rate due to requiring both ends of the connection be working.  This quadratic dependence implies low initial value so we expect  Narrowcast and Mobile-assisted contact tracing to provide initial value in adoption while privacy-sensitive mobile tracing provides substantial additional value once adoption rates are high.
 
 We note that there are an increasing number of concurrent contact tracing protocols being developed -- see in particular Section~\ref{sec:comparisons} for a discussion of solutions based on proximity based tracing (as in Figure~\ref{fig:bubble}).  In particular, there are multiple concurrent approaches using proximity based signaling; our approach has certain advantageous properties, as it is particularly simple and requires very little data transfer.
 
One point to emphasize is that, with this large number of emerging solutions, it is often difficult for the user to interpret what ``privacy preserving'' means in many of these protocols\footnote{In fact, due to re-identification risks, there is a strong case to be made that the terminology of ``privacy preserving'' is ill-suited to this context.}. One additional goal in providing the concrete protocols herein is to  have a broader discussion of both privacy-sensitivity and security, along with a transparent discussion of the associated re-identification risks ---  the act itself of alerting a user to being at risk provides de-anonymizing information, as we discuss shortly.

 \emph{From a civil liberties standpoint, the privacy guarantees these protocols ensure are designed to be consistent with the disclosures already extant in contract tracing methods done by public health services} (where some information from a positive tested citizen is revealed to other at risk citizens). In short, we seek to empower public health services, while maintaining civil liberties.
 
We also note that these contact tracing solutions are not meant to replace conventional contact tracing strategies employed by public health organizations; not everyone has phones, and not everyone that has a phone will use this app. Therefore, it is still critical to leverage conventional approaches, along with the approaches outlined in this paper. In fact, two of our protocols are designed for assisting public health organizations (and are designed with input from public health organizations).

\section{FAQ: Privacy, Security, and Re-Identification} \label{sect:FAQ}
Throughout, we refer to an \emph{at risk} individual as one who has been in contact with an individual who has tested as positive for COVID-19 (under criteria as defined by public health programs, e.g., ``within 6 feet for over 10 minutes'').

\begin{figure}
\begin{center}
\includegraphics[scale=0.6]{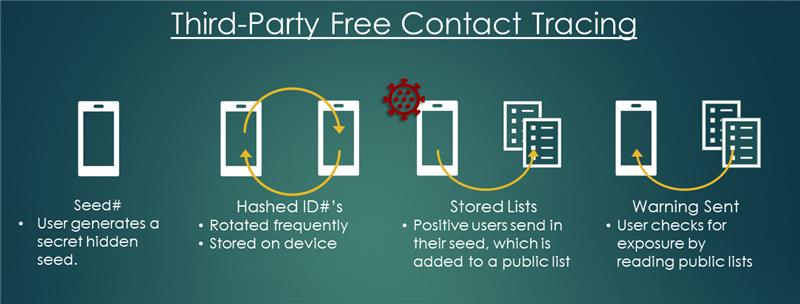}
\end{center}
\label{fig:cont} 
\caption{PACT Tracing Protocol. First, a user generates a random seed, which they treat as private information.
Then all users broadcast random-looking signals to users in their proximity via Bluetooth and, concurrently, all users also record all the signals they hear being broadcast by other users in their proximity.  Each person's broadcasts (their ``pseudonyms'') are a function of their private seed, and they change these broadcasted pseudonyms periodically (e.g. every minute).
Whenever a user tests positive, the positive user then can voluntarily publish, on a public server, information which enables the reconstruction of all the signals they have broadcasted to others during the infection window (precisely, they publish their private seed, and, using the seed, any other user can figure out what pseudonyms the positive user has previously broadcasted). Now, any other user can determine whether they are at risk by checking whether the signals they heard are published on the server.
 Note that the ``public lists'' can be either lists from hospitals, which have confirmed seeds from positive users, or they can be self-reports (see Section~\ref{sect:lab}). Credit: M Eifler.
}
\end{figure}

Before we start this discussion, it is helpful to consider one principle which the proposed protocols respect:
 ``If you do not report as being positive, then no information of yours will leave your phone." From a more technical standpoint, the statement that is consistent with our protocols is:
\begin{center}
 \emph{If you do not report as being positive, then only random  (``pseudonymized'') signals\\
 are permitted to be broadcast from your phone.} 
\end{center}
These random broadcasts are what allows proximity based tracing; see Figure~\ref{fig:cont} for a description of the mobile tracing protocol.
It is worthwhile to note that this principle is consistent, in spirit, with conventional contract tracing approaches, where only positively tested individuals reveal information to the public health authorities.  With the above principle, the discussion at hand largely focuses on what can be inferred when a positive disclosure occurs along with how a malicious party can impact the system.

We focus the discussion on the ``mobile tracing" protocol for the following reasons:  ``Narrowcasting" allows people to listen for events in their region, so it can viewed as a one way messaging system. For ``mobile-assisted interviews," all the data remains on the user's device, except for what they voluntarily reveal to public health authorities in order to enable contact tracing.

All the claims are consequences of basic security properties that can formally be proved about the protocol, and in particular, about the cryptographic mechanism generating these random-looking signals. 

\subsection{Confidentiality, Re-Identification, and Inferential Risks}

We start first with what private information is protected and what is shared voluntarily, following disclosure and consent.  The inferential risk is due to that the alert itself is correlated with other information, from which a user could deduce de-anonymizing information.

\begin{enumerate}[leftmargin=*]
\item If I tested positive and I voluntarily disclose this information, what does the protocol reveal to others?

Any other citizen who uses a mobile application following this protocol who has been at risk is notified. In some versions the time(s) that the exposure(s) occurred may be shared.  In the basic mobile tracing system that we envision, beyond exposure to specific individuals, no information is revealed to any other citizens or entities (authorities, insurance companies, etc). 

It is also worthwhile noting that, if you are negative, then the protocol does not directly transmit \emph{any} of your private information to any public database or any other third party; the protocol does transmit random (``pseudonymized") signals that your phone broadcasts. 

\item \textbf{Re-Identification and Inferential Risks.} Can a positive citizen's identity, who chooses to report being positive, be \emph{inferred} by others? 

Identification is possible and is a risk to volunteers who would prefer to remain de-identified.  Preventing proximity-based identification of this sort is not possible to avoid in any protocol, even in manual contact tracing as done by public health services, simply because the exposure alert may contain information that is correlated with identifying information.
For example, an individual who had been in close proximity to only one person over the last two weeks can infer the identity of this positively tested individual. 
However, the positive's identity will never be explicitly broadcast. In fact, identities are not even stored in the dataset: It is only the positive person's random broadcasts that are stored.

\item \textbf{Mitigating Re-identification.} Can the app be designed so as to mitigate re-identification risks to average users?

While the protocol itself allows a sophisticated user, who is at risk, to learn the time at which the exposure occurred, the app itself can be designed to mitigate the risk. For example, in the app design, the re-identification risk could be mitigated by only informing the user that they are at risk, or the app could only provide the rough time of day at which the exposure occurred.  This is a mild form of mitigation, which a malicious or sophisticated user could try to circumvent.

\end{enumerate}

\subsection{Attacks}

 We now directly address questions about the potential for malicious hackers, governments, or organizations to compromise the system. In some cases, cryptographically secure procedures can prevent certain attacks, and, in other cases, malicious disclosure of information is prevented because the protocol stores no data outside of your device by default. Only cryptographically secure data from positively confirmed individuals is  stored outside of devices.

\begin{enumerate}[leftmargin=*]

\item \textbf{Integrity Attacks.} If you are negative, can a malicious citizen listen to your phone's broadcasts and, then report positive pretending to be you? 

   No, this is not possible, provided you keep your initial seed private (see Figure~\ref{fig:cont}). Furthermore, even if the malicious party records all Bluetooth signals going into and out of your phone, this is not possible.
   
   This attack is important to avoid, since, suppose a malicious entity observes all Bluetooth signals sent from your phone. Then, you would not want this entity to report you as positive. This attack is not possible as the seed uniquely identifies your broadcasts and remains unknown to the attacker, unless the attacker is able to successfully break the underlying cryptographic mechanism, which is unlikely to be possible.

  \item \textbf{Inferential Attacks.} Can a positive citizen's location, who chooses to report being positive, be inferred by others?
  
    It is possible for a malicious party to simultaneously record broadcasts at multiple different locations, including those that the positive citizen visited. Using these recordings, the malicious party could infer where the  positive citizen was. The times at which the citizen visited these locations can also be inferred.

    \item \textbf{Replay and Reliability Attacks.}  If a citizen is alerted to be at risk, is it possible the citizen was not in the proximity of a positive individual?  
    
    There are a few unlikely attacks that can trigger a false alert. 
    One is a {\em replay attack}. For example, suppose a malicious group of multiple individuals colludes to try and pretend to be a single individual; precisely, suppose they all use the same private seed (see Figure~\ref{fig:cont}).  Then if only one of these malicious individuals makes a positive report, then multiple people can be alerted, even if those people were not in the proximity of the person who made the positive report.  The protocol incorporates several measures to make such attacks as difficult as possible.

    \item \textbf{Physical Attacks.} What information is leaked if a citizen's device is compromised by a hacker, stolen, or physically seized by an authority?
    
    Generally, existing mechanisms protect access to the storage of a phone. Should these mechanisms fail, the device only stores enough information to reconstruct the signals broadcast over a period of time prior to the compromise which amounts  to the length of the infection window (i.e., two weeks), in addition to collected signals. This enables some additional inference attacks. It is not possible to learn whether the user has ever reported positive. 

\end{enumerate}

\subsection{Lab-Based and Self-Confirmed Reporting; Reliability Concerns} \label{sect:lab}
Given that we would like the protocol to be of use to different states and countries, we seek an approach which allows for both security in reporting and for flexibility from the app designer in regions where it may make sense to  consider reports which are self-confirmed positives tests or self-confirmed symptoms.

\begin{enumerate}[leftmargin=*]

\item \textbf{Reporting.} Does the protocol support both medically confirmed positive tests and self-confirmed positives tests? 

Yes, it supports both. The uploaded files contain signatures from the uploading party (i.e. from a hospital lab or from any app following the protocol). This permits an app designer the freedom to use information from health systems and information from individuals in possibly different manners. In less developed nations, it may be helpful to permit the app designer to allow for reports based on less reliable signatures. 

\item \textbf{Reliability.} How will the protocol handle issues of false positives and false negatives, with regards to alerting? What about cases when users don't have (or use their) mobile phones?

The protocol does not explicitly address this, but a deployment requires both thoughtful app design and responsbile communication with the public.

With regards to the former, the false positive and false negative rates have to be taken into account when determining how to make at risk reports. More generally, estimates of the probabilities can be helpful to a user (or an otherwise interpretable report); such reports can be particularly relevant for those in high risk categories (such as the elderly and immuno-compromised individuals).  

Furthermore, not everyone has a smartphone,  and not everyone with a smartphone will use this app. Thus, users of this app -- if they have not received any notification of exposure with COVID-19 positive cases -- should not assume that they have not been around such positive cases. This means, for example, that they should still be cautious and follow all appropriate current public health guidelines, even if the app has not alerted them to possible COVID-19 exposure. This is particularly important until there is sufficient penetration of the app in any local population.

\end{enumerate}

\subsection{Other Threats (Exogenous to the Protocols)}

We now list threats that are outside of the scope of the protocol, yet important to consider. Care should be taken to address these concerns:
\begin{itemize}
    \item {\bf Trusted Communication.} Communication between users and servers must be protected using standard mechanisms (i.e., the TLS protocol~\cite{TLS}).
    \item {\bf Spurious Entries.} Self-reporting allows a malicious user to report themselves positive when they are not, and generally may allow several fake reports (i.e. a flooding attack). Mitigation techniques should be introduced to reduce the risk of such attacks.  
    \item {\bf Invalid Authentication.} Positive reports should be validated using digital signatures, e.g., by healthcare providers. This requires appropriate public-key infrastructure to be in place. Additional vulnerabilities related to misuse or misconfiguration of this infrastructure can affect reliability of positive reports.
 \item {\bf Implementation Issues.} Implementation aspects may weaken some of our claims, and need to be addressed. For example, signals we send over Bluetooth as part of our protocol may be correlated with other signals which de-anonymize the user.

\end{itemize}

\section{Protocols} 
We now provide an overview of the three functionalities of PACT.

\subsection{Privacy-sensitive Mobile Tracing} \label{sect:Bluetooth}

This section describes and discusses a privacy-sensitive mobile tracing protocol. Our protocol follows a pattern wherein users exchange IDs via Bluetooth communication.   If a user is both infected (we refer to such users as {\em positive}, and otherwise as {\em negative}) and willing to warn others who may have been at risk via proximity to the user, then de-identified information is uploaded to a server to warn other users of potential exposure. The approach has been followed by a number of similar protocols -- we describe the differences with some of them in Section~\ref{sec:comparisons}. In Appendix~\ref{sect:alternative}, we discuss an alternative approach which may offer some efficiency and privacy advantages, at the cost of relying on signatures as opposed to hash functions.

\subsubsection{Protocol description} 

Low-level technical details are omitted, e.g., how values are broadcast. Further, it is assumed the communication between users and the server is protected using the Transport Layer Security (TLS)  protocol. We first describe a variant of the protocol {\em without} entry validation, and discuss how to easily extend it to validate entries below.

\newcommand{\id}{\mathrm{id}}
\newcommand{\dt}{\mathrm{dt}}

\begin{quote}
\begin{itemize}
    \item {\bf Parameters.} We fix an understood time unit $\dt$ and define $\Delta$ such that $\Delta \cdot \dt$ equals the infection window. (Typically, this would be two weeks.) We also fix the bit length $n$ of the identifiers. (Typically, $n = 128$.) We also use a function $G: \{0,1\}^n \to \{0,1\}^{2n}$ which is assumed to be a secure cryptographic {\em pseudorandom generator} (PRG).\footnote{This means that its output, under a random input, is indistinguishable from a random string to a computationally bounded adversary.} If $n = 128$, we can use $G(x) = \textrm{SHA-256}(x)$.
    \item {\bf Pseudorandom ID Generation.} Every user broadcasts a sequence of IDs $\id_1, \id_2, \ldots \in \{0,1\}^n$.  The $i$-th $\id_i$ is broadcast at any time in the window $[t_0 + \dt\cdot (i-1), t_0 + \dt\cdot i[$, where $t_0$ is the start time. To generate these IDs, the user initially samples a random $n$-bit seed $S_0$, and then computes
\begin{displaymath}
(S_i, \id_i) \gets G(S_{i-1}) \;
\end{displaymath}
for $i = 1,2, \ldots$.
After $i$ time units, the user only stores $S^* \gets S_{\max\{i-\Delta,0\}}$, the time $t^*$ at which $S^*$ was generated, the current $S_i$, and the time $t_i$ at which $S_i$ was generated. Note that if the device was powered off or the application disabled, we need to advance to the appropriate $S_i$. 

\item {\bf Pseudorandom ID Collection.} For every $\id$ broadcast by a device in its proximity at time $t$, a user stores a pair $(\id, t)$ in its local storage $\mathcal{S}$.
\item {\bf Reporting.} To report a positive test, the user uploads $(S^*, t_{start}=t^*, t_{end}=t_i)$ to the server, which appends it to a public list $\mathcal{L}$. The server checks that $t_{start}$ and $t_{end}$ are reasonable before accepting the entry. Once reported, the user erases its memory and restarts the pseudorandom ID generation procedure.
\item {\bf Checking Exposure.} A user downloads $\mathcal{L}$ from the server (or the latest portion of it). For every entry $(S^*,t_{start}, t_{end})$ in $\mathcal{L}$, it generates the sequence of IDs $\id^*_1, \ldots, \id^*_{\Delta}$ starting from $S^*$, as well as  estimates $t_i^*$ of the time at which each $\id_i^*$ was initially broadcast. If $\mathcal{S}$ contains $(\id_i^*, t)$ for some $i \in \{1, \ldots, \Delta\}$ such that $t$ and $t_i^*$ are sufficiently close, the user is alerted of potential exposure.
\end{itemize}
\end{quote}
{\bf Setting Delays.} To prevent replay attacks, an entry $(S^*,t_{start}, t_{end})$ should be published with a slight delay. This is to prevent an $\id^*_{\Delta}$ generated from $S^*$ being recognized as a potential exposure by any user if immediately rebroadcast by a malicious party.

\noindent {\bf Entry Validation.} Entries can (and should) be validated by attaching a signature $\sigma$ on $(S^*, t_{start}, t_{end})$ when reporting, as well as (optionally) a certificate to validate this signature. An entry thus has form $(S^*, t_{start}, t_{end}, \sigma, cert)$. Entries can be validated by multiple entities, by simply re-uploading them with a new signature.

A range of designs and policies are supported by this approach. Upon an initial update, a (weakly secure) signature with an app-specific key could be attached for self-reporting. This signature does not provide any real security (as we cannot guarantee that an app-specific signing key remains secret), but can be helpful to offer improved functionality.  
Third-parties (like health-care providers) can re-upload an entry with their signature after validation. An app can adopt different policies on how to display a potential exposure depending on how it is validated.

We also do not specify here the infrastructure required to establish the validity of certificates, or how a user interacts with a validating party, as this is outside the scope of this description. 

\noindent {\bf Fixed-Length Sequences of IDs.}
As stated, during the first $\Delta-1$ time units a user will have generated a sequence of fewer than $\Delta$ IDs.
During this time, the number of IDs the user has generated from its current $S^\ast$ is determined by how long ago the user started the current pseudorandom ID generation procedure (either when they initially started using the protocol or when they last submitted a report).   
This may be undesirable information to reveal to a party that gains access to the sequence of IDs (e.g. if the user submits a report or if the party gains physical access to the user's device). 
So to avoid revealing this information, a user may optionally iterate to $S_{\Delta}$ and use $\id_{\Delta}$ as the first ID they broadcast when starting or restarting the pseudorandom ID generation procedure.

{\bf Synchronized Updates.}
Suppose a user updates their seed every $\dt$ amount of time after whenever they happened to originally start the ID generation process.
Then it may be possible to correlate two IDs of a user by noticing that the times at which the IDs were initially broadcast were separated in time by a multiple of $\dt$.
To mitigate this it would be beneficial to have an agreed schedule of when all users update their seed.
For example, if $\dt$ is 15 minutes then it might be agreed that everyone should update their seed at midnight UTC, followed by 12:15, 12:30, and so forth.


\subsubsection{Privacy and Security Properties}

Privacy and integrity properties of the protocol follow from the following two propositions. (Their proofs are omitted and follow from standard techniques.) In the following discussion, it is convenient to refer to an ID value $\id_i$ output by a user as {\em unreported} if it is not within the $\Delta$ $\id$'s generated by a seed the user has reported to the server. 

\begin{proposition}[Pseudorandomness]
   All unreported IDs are pseudorandom, i.e., no observer (different than the user) can distinguish them from random looking strings (independent from the state of the user) without compromising the security of $G$.
\end{proposition}

\begin{proposition}[One-wayness]
   No attacker can produce a seed $S$ which generates a sequence of $\Delta$ IDs that include an unreported ID generated by an honest user (not controlled by the adversary) without compromising the security of $G$.
\end{proposition}

To discuss the consequences of these properties on privacy and integrity, let us refer to users as either ``positive'' or ``negative'' depending on whether they decided to report as positive, by uploading their seed to the server, or not. 
\begin{itemize}
    \item {\em Privacy for negative users.}  By the pseudorandomness property, a negative user $u$ only broadcasts pseudorandom IDs. These IDs cannot be linked without knowledge of the internal state of $u$. This privacy guarantee improves with the frequency of updating the seed $S_i$ -- ideally, if a different $\id_i$ is broadcast each time, no linking is possible. This however results in less efficient checking for exposure by negative users.\footnote{In a Bluetooth implementation, one needs to additionally ensure that each different $\id_i$ is broadcast with a different UUID to prevent linking.}
    \item {\em Privacy for positive users.} Upon reporting positive, the last $\Delta$ IDs generated by the positive user {\em can} be linked. (We discuss what this means below, and possible mitigation approaches.) However, by pseudorandomness, this is only true for the IDs generated within the infection window. Older IDs and {\em newer} IDs cannot be linked with those in the infection window, and with each other. Therefore, a positive user has the same guarantees as a negative user outside of the reported infection window.
    \item {\em Integrity guarantees.} It is infeasible for an attacker to upload to the server a value $S^*$ which generates an unreported ID that equals one generated by another user. This prevents the attacker from misreporting IDs of otherwise negative users and erroneously alerting their contacts.  
\end{itemize}

{\bf Timing Information and Replay Attacks.} 
The timestamping is necessary to prevent {\em replay attacks}. In particular, we are concerned by adversaries rebroadcasting IDs of legitimate users (to be tested positive) outside the range of their devices. This may create a high number of false exposures to be reported. 

An attack we cannot prevent is the following {\em relay attack}: An attacker captures an ID of an honest user at location A, sends it over the Internet to location B, where it is re-broadcast.  However, as soon as there is sufficient delay, the attack is prevented by maintaining sufficiently accurate timing information. (One can envision several accuracy compromises in the implementation, which we do not discuss here.)

{\bf Strong Integrity.} Our integrity property does not prevent a malicious user from reporting a seed $\widetilde{S}^*$ generating an ID which has been already reported. Given an entry with seed $S^*$, the attacker just chooses (for example) $\widetilde{S}^*$ as the first half of $G(S^*)$. The threat of such attacks does not appear significant. However, they could be prevented with a less lightweight protocol, as we explain next. We refer to the resulting security guarantee as {\em strong integrity}. 

Each user generates a signing/verification-key pair $(sk, vk)$ along with the initial seed. Then, we include $vk$ in the ID generation process, in particular let $(S_i, \id_i) \gets G(S_{i-1}, vk)$. An entry now consists of $(S^*, t_{start}, t_{end}, vk, \sigma)$, where $\sigma$ is a signature (with signing key $sk$) on $(S^*, t_{start}, t_{end}, vk)$. Entries with invalid signatures are ignored. (This imposes slightly stronger assumptions on $G$ -- pseudorandomness under related seeds sharing part of the input and binding of $vk$ to $S_i$.)

The CEN protocol, discussed in Section~\ref{sec:comparisons}, is the only one that targets strong integrity, though their initial implementation failed to fully achieve it. (The issue has been fixed after our report.)

\subsubsection{Inference from Positive IDs} 

One explicit compromise we take is that {\em IDs of a positive user can be linked within the infection window}, and that the start and end time of the infection window is known. For example, an adversary collecting IDs at several locations can detect that the same positive user has visited several locations at which it collects broadcast identifiers. This can be abused for surveillance purposes, but arguably, surveillance itself could be achieved by other methods. The most problematic aspect is the linking of this individual with the fact that they are positive. 

A natural approach to avoid linking, as in~\cite{BU}, is for the the server to only expose the IDs, rather than a seed from which they are computed. However, this does not make them unlinkable. Imagine, at an extreme, that the storage on the server is append only (which is a realistic assumption). Then, the IDs belonging to the same user are stored sequentially. One can obfuscate this leakage of information in several ways, for example by having the server buffer a certain amount of new IDs, and shuffle them before release. Nonetheless, the actual privacy improvement is hard to assess without a good statistical model of upload frequency. This also increases the latency of the system which directly harms its public health value.

A user could also learn at which time the exposure took place, and hence infer the identity of the positive user from other available information. We stress that the application can and should refuse to display the time of potential exposure -- thus preventing a ``casual attacker'' from learning timing information. However, a malicious app can always remember at which time an ID has been seen.

\subsection{Mobile-Assisted Contact Tracing Interviews} 
Contact tracing interviews are laborious and often miss important events due to the limitations of human memory.  Our plan to assist here is to provide information to the end user that can (with consent) be shared with a public health organization charged with performing contact tracing interviews.  This is not an exposure of the entire observational log, but rather an extract of the information which is requested in a standard contact tracing interview. We have been working with healthcare teams from Boston and the University of Washington on formats and content of information that are traditionally sought by public health agencies.  Ideally, such extraction can be done working with the user before a contact tracing interview even occurs to speed the process.

\subsection{Narrowcasting}

Healthcare authorities from NYC have informed us that they would love to have the ability to make public service announcements which are highly tailored to a location or to a subset of people who may have been in a certain region during specific periods of time.  This capability can be enabled with a public server supporting (area x time,message) pairs.  Here ``area" is a location, a radius (minimum 10 meters), a beginning time and an ending time.  Only announcements from recognized public health authorities are allowed.  

Anyone can manually query the public server to determine if there are messages potentially relevant to them per their locations and dwells at the locations over a period of time. However, simple automation can be extremely helpful as phones can listen in and alert based on filters that are dynamically set up based on privately-held locations and activities. Upon downloading (area x time, message) pairs a phone app (for example) can automatically check whether the message is relevant to the user.  If it is relevant, a message is relayed to the device owner.

Querying the public server provides \emph{no} information to the server through the protocol itself, because only a simple copy is required.

\section{Alternative Approaches}\label{sec:alt}

We discuss some alternative approaches to mobile tracing. Some of these are expected to be adopted in existing and future contact-tracing proposals, and we discuss them here. 

Hart et al.~\cite{vihart} provides a useful high-level understanding of the issues involved in contact tracing. They discuss, among other topics, the value of using digital technology to scale contract tracing and the trade-offs between different classes of solutions.

\subsection{Reporting collected IDs}
\label{sec:dual}

PACT users upload their locally {\em generated} IDs upon a positive report. An alternative is to upload {\em collected} IDs of potentially at risk users. This approach (which we refer to as the {\em dual} approach) has at least one clear security disadvantage and one mild privacy advantage over PACT. (The latter is only true if the system is carefully implemented, as we explain below.)

{\bf Disadvantages: Reliability and Integrity Attacks.} In the dual approach, a malicious user cannot be prevented from very easily reporting a very large number of IDs which were not generated by users in physical proximity. These IDs could have been collected by colluding parties elsewhere, at any time before the report. Such attacks can seriously hurt the reliability of the system. In PACT, to achieve a similar effect, the attacker needs to (almost) simultaneously broadcast the same ID in direct proximity of all individuals who should be falsely alerted to be potentially at risk. 

PACT ensures integrity of positive reporting by exhibiting a seed generating these IDs, known only to the reporter. A user $u$ cannot frame another negative user $u'$ as a positive user by including an ID generated by $u'$. In the dual approach, user $u'$ could be framed for example by uploading IDs that have been broadcast in their surroundings.

{\bf Advantage: Improved Temporal Ambiguity.} Both in the dual approach and in PACT-like designs, a user at risk can de-anonymize a positive user from the time at which the matching ID was generated/collected, and other contextual information (e.g., a surveillance video). 

The dual approach offers a mitigation to this using {\em re-randomization} of IDs. We explain one approach~\cite{yael}. Let $\mathbb{G}$ be a prime-order cyclic group with generator $g$ (instantiated via a suitable elliptic curve). 
\begin{enumerate}
    \item Each user $u$ chooses a secret key $s_u$ as a random element in $\mathbb{Z}_p$.
    \item Each broadcast ID takes the form $\id_i = (g^{r_i}, g^{r_i s_u})$, where $r_1, r_2, \ldots$ are random elements of $\mathbb{Z}_p$.
    \item To upload an ID with form $\id = (x, y)$ with a report, a positive user uploads instead a re-randomized version $\id' = (x^r, y^r)$, where $r$ is a fresh random value from $\mathbb{Z}_p$.
    \item To determine whether they are at risk, user $u$ checks whether an ID of the form $\id = (x,y)$ such that $y = x^{s_u}$ is stored on the server.
\end{enumerate}
Under a standard cryptographic assumption -- the so-called {\em Decisional Diffie-Hellman} (DDH) assumption -- the IDs are pseudorandom. Further, a negative user who learns they are at risk cannot tell which one of the IDs they broadcast has been reported, as long as the reporting user re-randomized them and all IDs have been generated using the same $s_u$. Note that incorrect randomization only hurts the positive user.

Crucially, however, the privacy benefit inherently relies on each user $u$ re-using the same $s_u$, and we cannot force a malicious user to comply. For example, to track movements of positive users, a surveillance entity can generate IDs at different locations with form $(x,y)$ where $y = x^{s_L}$ and $s_L$ depends on the location $L$. Identifiers on the server with form $(x, x^{s_L})$ can then be traced back to location $L$. A functionally equivalent attack is in fact more expensive against PACT, as this would require storing all IDs of users broadcast at location $L$.

\subsection{Centralized (Trusted Third-Party) Approaches}\label{sect:centr}

We discuss an alternative {\em centralized} approach here, which relies on a trusted third party (TTP), typically an agency of a government. Such a solution requires an initial {\em registration phase} with the TTP, where each user subscribes to the service. Moreover, the protocol operates as follows:
\begin{enumerate}    
\item Users broadcast random-looking IDs and gather IDs collected in their proximity. 
\item Upon a positive test, a user reports to the TTP all of the IDs collected in their proximity during the relevant infection window. The TTP then alerts the users who generated these IDs, who are now at risk.
\end{enumerate}
In order for the TTP to alert potentially at risk users, it needs to be able to identify the owners of these identifiers. There a few technical solutions to this problem. 
\begin{itemize}
\item One option is to have the TTP generate all IDs which are used by the users - this requires either storing them or (in case only seeds generating them are stored) a very expensive check to identity at risk users.
\item A more efficient alternative for the TTP (but with larger identifiers) goes as follows. The trusted third-party generates a public-key/secret-key pair $(sk, pk)$, making $pk$ public. It also gives a unique token $\tau_u$ to each user $u$ upon registration, which it remembers. Then, the $i$-th ID of user $u$ is $\id_i = \mathrm{Enc}(pk,\tau_u)$. (Note that encryption is randomized here, so every $\id_i$ appears independent from prior ones.) The TTP can then efficiently identify the user who generated $\id_i$ by decrypting it. 
\end{itemize}

{\bf Privacy Considerations.} Such a centralized solution offers better privacy against attackers who do not collude with the TTP - in particular, only pseudorandom identifiers are broadcast all times. Moreover, at risk individuals only learn that one of the IDs they collected belongs to a positive individual. A -risk users can still collude, learning some information from the time of being reported at risk, and correlate identifiers belonging to the same positive user, but this is harder.  

The biggest drawback of this solution, however, is the high degree of trust on the TTP. For example:
\begin{itemize}
    \item The TTP learns the identities of all at risk users who have been in proximity of the positive subject.  
    \item The TTP can, at any time and independently of any actual report, learn the identity of the user $u$ who broadcasts a particular ID, or at least link them to their token $\tau_u$. This could be easily exploited for surveillance of users adopting the service.
\end{itemize}

{\bf Security Consideration.} As in the dual approaches described above, it is trivial for a malicious party identifying as honest to report valid identifiers of other users (which may have been collected in a distributed fashion) to erroneously alert them as being at risk. Replay attacks can be mitigated by encrypting extra meta-data along with $\tau_u$ (e.g., a timestamp), but this would make IDs even longer.

If the TTP is malicious it can target specific users to falsely claim they are at risk or to refrain from informing them when they actually are at risk.

\subsection{Absolute-Location--Centric Mobile Tracing Methods}\label{sect:GPS}
It is also possible to design protocols based on the sensing of absolute locations (GPS, and GPS extended with dead reckoning, wifi, other signals per current localization methods) consistent with ``If you do not report as being positive, then no information of yours will leave your phone'' (see Section~\ref{sect:FAQ}).  For example, a system could upload location traces of positives (cryptographically, in a secure manner), and then negative users, whose traces are stored on their phones could intersect their traces with the positive traces to check for exposure. This could potentially be done with stronger cryptographic methods to limit the exposure of information about these traces to negative users; one could think of this as a more general version of {\em private-set intersection} (PSI)~\cite{PSI,PSI2,PSI3}. However, such solutions would still reveal traces of positives to a server.  

There are two reasons why we do not focus on the details of such an approach here:
\begin{itemize}
    \item Current localization technologies are not as accurate as the use of Bluetooth-based proximity detection, and may not be accurate enough to be consistent with medically suggested definitions for exposure.
    \item Approaches employing the sensing and collection of absolute location information would need to rely more heavily on cryptographic protocols to keep the positive users traces secure.
\end{itemize}
However, this is an approach worth keeping in mind as an alternative, per assessments of achievable accuracies and relevance of the latter accuracies for public health applications.

\section{Related Efforts}
\label{sec:alternatives}
\label{sec:comparisons}

There are an increasing number of contact tracing applications being created with different protocols.
We will briefly discuss a few of these and how their mobile tracing protocols compare with the approaches described in Section \ref{sect:Bluetooth} and \ref{sec:alt}.

\subsection{Similar Bluetooth-Based Efforts}
The privacy-sensitive mobile tracing protocols proposed by
\href{https://www.coepi.org/}{CoEpi}~\cite{coepi},
\href{https://www.covid-watch.org/}{CovidWatch}~\cite{covidwatch},
as well as \href{https://github.com/DP-3T/documents}{$\mathrm{DP^3T}$}~\cite{dp3t}, 
have a similar structure to our proposed protocol.
We briefly describe the technical differences between all of these protocols and discuss the implications of these differences.

Similar to our proposed protocol, these are based on producing pseudorandom IDs by iteratively applying a PRG $G$ to a seed.
CoEpi and CovidWatch use the Contact Event Numbers (CEN) protocol, in which the initial seed is derived from a digital signature signing key $rak$ and $G$ is constructed from two hash functions (which during each iteration incorporate an encoding of the number of iterations done so far and the verification key $rvk$ which matches $rak$).
Another proposal is the $\mathrm{DP^3T}$~\cite{dp3t} protocol, in which $G$ is constructed from a hash function, a PRF, and another PRG.
The latter PRG is used so that a single iteration of $G$ produces all the IDs needed for a day.
These IDs are used in a random order throughout the day.
Both of these (under appropriate cryptographic assumptions) achieve the same sort of pseudorandomness and one-wayness properties as our protocol.

The incorporation of $rvk$ into $G$ with CEN is intended to provide strong integrity and allow a reporting user to include a memo with their report that is cryptographically bound to the report. Two ideas for what such a memo might include are a summary of the user's self-reported symptoms (CoEpi) or an attestation from a third party verifying that the user tested positive (CovidWatch). Because a counter of how many times the seed has been updated is incorporated into $G$, a report must specify the corresponding counters. This leaks how long ago the user generated the initial seed, which could potentially be correlated with identifying information about the user (e.g., when they initially downloaded the app). 

An earlier version of CEN incorrectly bound the digital signature key to the identifiers in a report.
Suppose an honest user has submitted a report for $id_{j}$ through $id_{j'}$ (for $j<j'$) with a user chosen memo. 
Given this report, an attacker could create their own report that verifies as valid, but includes the honest user's $id_{i}$ for some $i$ between $j$ and $j'$ together with a memo of the attacker's choosing.
A fix was proposed after we contacted the team behind the CEN protocol.

The random order of a user's IDs for a day by $\mathrm{DP^3T}$ is intended to make it difficult for an at risk individual to identify specifically when they were at risk (and thus potentially, by whom they were exposed).
A protocol cannot hope to hide this sort of timing information from an attacker that chooses to record the time when they received every ID they see; this serves instead as a mitigation against a casual attacker using an app that does not store this sort of timing information.  
In our protocol and CEN, information about the exposure time is not intended to be as hidden at the protocol. In our protocol the time an ID was used is even included as part of a report and used to prevent replay attacks, as discussed earlier.
CEN does not use timing information to prevent replay attacks, but considers that an app may choose to give users precise information about where they were exposed (so the user can reason about how likely this potential exposure was to be an actual exposure).

A similar protocol idea was presented in~\cite{BU}. 
It differs from the aforementioned proposals in that 
individual IDs are uploaded to the server, rather than a seed generating them (leading to increased bandwidth and storage).
Alternatives using bloom filters to reduce storage are discussed, but these inherently decrease the reliability of the system. $\mathrm{DP^3T}$ also recently included a similar protocol as an additional option, using cuckoo filters in place of bloom filters.

\subsection{Centralized Example}
The \href{https://www.tracetogether.gov.sg/}{TraceTogether}~\cite{tracetogether} app is currently deployed in Singapore.
It uses the BlueTrace protocol designed by at team at the Government Technology Agency of Singapore.
This protocol is closely related to the encryption-based technique discussed in Section \ref{sect:centr}.

\subsection{Absolute-Location--Centric Example}
The \href{http://safepaths.mit.edu/}{Private Kit: Safe Paths} app~\cite{safepaths,safepaths2} intends to use an absolute-location--centric approach to mobile tracing.
They intend to mitigate some of the downsides discussed in Section \ref{sect:GPS} by reported location traces of positive users to be partially redacted.
It is unclear what methodology they intend to use for deciding how to redact traces. The trade-off in this redaction process between how easily a positive user can be identified from their trace and how much information must be removed from it (decreasing its usefulness). 
They intend to use cryptographic protocols (likely based on \cite{safepathsmaybe}) to minimize the amount of information revealed about positive users' traces.

\subsection{Other Efforts}
A group of scientists at the Big Data Institute of Oxford University have proposed the use of a mobile contact-tracing app~\cite{oxford,oxford2} based on their analysis in~\cite{oxford3}.
The \href{https://nexttrace.org/}{NextTrace}~\cite{nexttrace} project aims to coordinate with COVID-19 testing labs and users, providing software to enable contact tracing.
The details of these proposals and the privacy protections they intend to provide are not publicly available.

The projects we refer to are only a small selection of the mobile contract-tracing efforts currently underway.
A more extensive listing of these projects is being maintained at~\cite{gdoc}, along with other information of interest to contract tracing.

\section{Discussion and Further Considerations}

\subsection{Interoperability of Different Protocols}

Most protocols like ours store a seed on a server, which is then used to deterministically generate a sequence of identifiers. Details differ in how exactly these sequences are generated (including the adopted cryptographic algorithms).  However, it appears relatively straightforward for apps to be modified to support all of these different sequence formats. A potential challenge is data from different protocol may provide different levels of protection (e.g., the lack of timing information may reduce the effectiveness against replay attacks). This difference in reliability may be surfaced via the user-interface.

In order to support multiple apps accessing servers for different services, it is important to adopt an interoperable format for entries to be stored on a server and possibly, to develop a common API.

\subsection{Ethics Considerations}
We acknowledge that ethical questions arise with contact tracing and in the development and adoption of any new technology. The question of how to balance what is revealed for the good of public health vs individual freedoms is one that is central to public health law. We iterate that privacy is already impacted by tracing practices. In some nations, positively tested citizens are required, either by public health policy or by law, 
to disclose aspects of their history. Such actions and laws frame multiple concerns about privacy and freedom, and bring up important questions.  The purpose of this document is lay out some of the technological capabilities, which supports broader discussion and debate about civil liberties and the risks that contact tracing can pose to civil liberties.

Another concern is accessibility to the service: not everyone has a phone (or will have the service installed). One consequence of this is that the quality of contract tracing in a certain population inherently depends on factors orthogonal to the technological aspects, which in turn raises important questions about fairness.

\subsection{Larger Considerations of Testing, Tracing, and Timeouts}
Tracing is one part of a conventional epidemic response strategy, based on Tests, Tracing, and Timeouts (TTT). Programs involving all three components are as follows:
\begin{itemize}
\item Test heavily for the virus.  South Korea ran over 20 tests per person found with the virus. 
\item Trace the recent physical contacts for anyone who tests positive.  South Korea conducted \emph{mobile contact tracing} using telecom information.
\item Timeout the virus by quarantining contacts until their immune system purges the virus, rendering them non-infectious.
\end{itemize}
The mobile tracing approach allows this strategy to be applied at a dramatically larger scale than only relying on human contact tracers.  

\subsection{Challenge of Wide-Scale Adoption}
This chain is only as strong as its weakest link.  Widespread testing is required and wide-scale adoption must occur. Furthermore, strategies must also be employed so that citizens takes steps to self-quarantine or seek testing (as indicated) when they are exposed. We cannot assume 100 percent usage of the application and concomitant enlistment in TTT programs.  Studies are needed of the efficacy of the sensitivity of the effectiveness of the approach to different levels of subscription in a population.

\subsection*{Acknowledgments}
The authors thank Yael Kalai for numerous helpful discussions, along with suggesting the protocol outlined in Section~\ref{sec:dual}. We also thank Ramesh Raskar and the Microsoft Cryptography group including Kristin Lauter, Kim Laine, Esha Ghosh, and Melissa Chase for private set intersection discussions related to locations.
We thank Edward Jezierski, Nicolas di Tada, Vi Hart, Ivan Evtimov, and Nirvan Tyagi for numerous helpful discussions. We also graciously thank M Eifler for designing all the figures.
Sham Kakade acknowledges funding from the Washington Research
Foundation for Innovation in Data-intensive Discovery, the ONR award
N00014-18-1-2247, NSF grants \#CCF-1637360 and \#CCF 1740551. Jacob Sunshine acknowledges
funding from NIH (K23DA046686) and NSF (1914873, 1812559). Stefano Tessaro acknowledges support from a Sloan Research Fellowship and from the NSF under grants CNS-1553758, CNS-1719146.

\bibliographystyle{abbrv}
\bibliography{main.bib}

\newpage
\appendix

\section*{Appendix}

\section{Issues around Practical Implementation}

A number of practical issues and details may arise with implementation.
\begin{enumerate}
    \item With regards to anonymity, if the protocol is implemented over the internet, then GeoIP lookups can be used to localize the query-maker to a varying extent.  People who really care about this could potentially query through an anonymization service.
    \item The narrowcast messages in particular may be best expressed through existing software map technology.   For example, we could imagine a map querying the server on behalf of users and displaying public health messages on the map.  
    \item The bandwidth and compute usage of a phone querying the full database may be too high.  To avoid this, it's reasonably easy to augment the protocol to allow users to query within a (still large) region.  We mention one such approach below.  
    \item Disjoint authorities.  Across the world, there may be many testing authorities which do not agree on a common infrastructure but which do wan to use the protocol.  This can be accommodated by enabling the phone app to connect to multiple servers. 
    \item The mobile proximity tracing does not directly inform public authorities who may be a contact.  However, it does provide some bulk information, simply due to the number of posted messages. 
\end{enumerate}

There are several ways to implement the server.  A simple approach, which works fine for not-too-many messages just uses a public GitHub repository.

A more complex approach supporting regional queries is defined next.  

\subsection{Regional Query Support}
Anyone can ask for a set of messages relevant to some region $R$ where $R$ is defined by a latitude/longitude range with messages after some timestamp.  More specific subscriptions can be constructed on the fly based on policies that consider a region $R$ and privately observed periods of time that an individual has spent in a region. Such scoped queries and messaging services that relay content based on location or on location and periods of time are a convenience to make computation and communication tractable.  The reference implementation uses regions greater in size than typical GeoIP tables.

To be specific, let's first define some concepts.
\begin{itemize}
    \item Region: A region consists of a latitude prefix, a longitude prefix, and the precision in each.  For example, New York which is at 40.71455 N, -74.00712 E can be coarsened to 40 N, -74 E with two digits of precision (the actual implementation would use bits).
    \item Time: A timestamp is specified in the number of seconds (as a 64 bit integer) since the January 1, 1970. 
    \item Location: A location consists of a full precision Latitude and Longitude
    \item Area: An area consists of a Location, a Radius, a beginning Time, and an ending Time.
    \item Bluetooth Message: A Bluetooth message consists of a fixed-length string of bytes.  It is used with the Bluetooth sensory log to discover if there is a match, which results in a warning that the user may have been in contact with an infected person.  
    \item Message: A message is a cryptographically signed string of bytes which is interpreted by the phone app. This is used for either a public health message (announced to the user if the sensory log matches) or a Bluetooth Message. 
\end{itemize}

With the above defined, there are two common queries that the server supports as well as an announcement mechanism.
\begin{itemize}
    \item GetMessages(Region, Time) returns all of the (Area, Message) pairs that the server has added since Time for the Region. The app can then check locally whether the Area intersects with the recorded sensory log of (Location,Time) pairs on the phone, and alert the user with the Message if so.   
    \item HowBig(Region, Time) returns the (approximate) number of bytes worth of messages that would be downloaded on a GetMessages call with the same arguments.  HowBig allows the phone app to control how much information it reveals to the server about locations/times of interest according to a bandwidth/privacy tradeoff.  For example, the phone could start with a very coarse region, specifying higher precision regions until the bandwidth required is acceptable, then invoke GetMessages.  (This functionality is designed to support controlled anonymity across widely varying population densities.)
    \item Announce(Area,Message) uploads an (Area, Message) pair for general distribution.  To prevent spamming, the signature of the message is checked against a whitelist defined with the server. 
\end{itemize}

\newcommand{\skey}{\mathrm{sk}}
\newcommand{\vkey}{\mathrm{vk}}
\newcommand{\Kg}{\mathsf{Kg}}
\newcommand{\Sign}{\mathsf{Sign}}
\newcommand{\Vrfy}{\mathsf{Vrfy}}
\newcommand{\gps}{\mathrm{coord}}

\section{Alternative Protocol}\label{sect:alternative}

We propose an alternative to the protocol in Section~\ref{sect:Bluetooth}. One main difference is that the server {\em cannot} generate the IDs broadcast by a positive user, and only stores a short verification key used to identify IDs broadcast by the positive user. While this does not prevent many of the inference scenarios we discussed above, this appears to be a desirable property. As we explain below, this protocol offers a different cost for checking exposure, which may be advantageous in some deployment scenarios.

This alternative approach inherently introduces risks of replay attacks which cannot be prevented by storing timestamps, because the server obtains no information about the times at which IDs have been broadcast. To overcome this, we build on top of a very recent approach of Pietrzak~\cite{EPRINT:Pietrzak20} for replay-attack protection. (Along similar lines, this can also be extended to relay-attack protection by including GPS coordinates, but we do not describe this variant here.)

\begin{quote}
\begin{itemize}
    \item {\bf Setup and Parameters.} We fix an understood time unit $\dt$. We make use of a \emph{digital signature scheme} specifying algorithms for key generation, signing, and verification, denoted $\Kg$, $\Sign$, and $\Vrfy$, respectively. We also use a hash function $H:\{0,1\}^n\times\{0,1\}^\ast\to\{0,1\}^n$. We can use $H(k,x)=\textrm{SHA256}(k||x)$ and Ed25519~\cite{bernstein,Ed25519} for the signature scheme.
    \item {\bf Pseudorandom ID Generation.} Every user broadcasts a sequence of IDs $\id_{1,d}, \id_{2,d}, \ldots$ during day $d$.  The $i$-th $\id_{i,d}$ is broadcast at any time in the window $[t_d+ \dt\cdot (i-1), t_d + \dt\cdot i[$, where $t_d$ is the time at the beginning of the day. To generate these IDs, the user runs $\Kg$ to generate a daily signing key / verification key pair $(\skey_{d}, \vkey_{d})$, and keeps both of them secret. They also determine the current time $t_i=t_d+ \dt\cdot (i-1)$. 
    Finally, the user samples $n$-bit random strings $R_i$ and $r_i$ and computes the identifier as 
		\begin{displaymath}
		\id_i = (\sigma_i, R_i, h_i), \;
		\end{displaymath}
		where $\sigma_i = \Sign(\skey_{d}, R_i|| h_i)$ and $h_i=H(r_i, t_i)$.
		They broadcast $(\id_i, r_i, t_i)$.
		When day $d$ ends the user deletes their signing key $\skey_d$. (The verification key $\vkey_d$ is {\em not} deleted, until an amount of time equal to the infection window has elapsed.)
	\item {\bf Pseudorandom ID Collection.} For every $(\id_i = (\sigma_i, R_i, h_i),r_i, t_i)$ broadcast by a device in their proximity, a user first checks if $t_i$ is sufficiently close to their current time and if $h_i = H(r_i, t_i)$. If so, they store $\id_i$ in their local storage $\mathcal{S}$.
	\item {\bf Reporting.} To report a positive test, the user uploads each of their recent $\vkey_d$ to the server, which appends them to a public list $\mathcal{L}$. Once reported, the user erases their memory and restarts the pseudorandom ID generation procedure.
	\item {\bf Checking Exposure.} A user downloads $\mathcal{L}$ from the server (or the latest portion of it). For every entry $\vkey$ in $\mathcal{L}$ and every entry $(\sigma, R, h)$ in $\mathcal{S}$, they run $\Vrfy(\vkey,\sigma,R|| h)$. If this returns true, the user is alerted of potential exposure.
\end{itemize}
\end{quote}
{\bf Efficiency Comparisons.} Let $\Delta$ be the number of IDs broadcast over the infection window. Let $S = |\mathcal{S}|$ be the size of the local storage. Let $L$ be the number of new verification keys a user downloads. To check exposure, the protocol from Section~\ref{sect:Bluetooth} roughly runs in time
\begin{displaymath}
    L \times \Delta \times \log(S) \times t_{G}  \;,
\end{displaymath}
where $t_{G}$ is the time needed to evaluate $G$. In contrast, for the protocol in this section, the time is
\begin{displaymath}
  L \times S \times t_{\Vrfy} \;,
\end{displaymath}
where $t_{\Vrfy}$ is the time to verify a signature. One should note that $t_{\Vrfy}$ is generally larger than $t_G$, but can still be quite fast. (For example, Ed25519 enables fast batch signature verification.)

Therefore, the usage of this scheme makes particular sense if a user does not collect many IDs, i.e., $S$ is small relative to $\Delta \cdot \log(S)$.


Another important point of comparison is how many bits need to be broadcast by the protocol.

{\bf Assumptions.} We require the following two standard properties for the hash function $H$:
\begin{itemize}
	\item \textbf{Pseudorandomess:} For any $x$ and a randomly chosen $r\in\{0,1\}^n$, the output $H(r,x)$ looks random to anyone that doesn't know $r$.
	\item \textbf{Collision resistance:} It is hard to find distinct inputs to $H$ that produce the same output.
\end{itemize}
Of our digital signature scheme we require the following three properties. The first is a standard property of digital signature schemes. The latter two are not commonly required of a digital signature scheme, so one needs to be careful when choosing a signature scheme to implement this protocol. We have verified that these properties are achieved by Ed25519 under reasonable cryptographic assumptions. 
\begin{itemize}
	\item \textbf{Unforgeability:} Given $\vkey$ and examples of $\sigma=\Sign(\skey,m)$ for attacker-chosen $m$, an attack cannot produce a new $(\sigma',m')$ for which $\Vrfy(\vkey, \sigma', m')$ returns true.
	\item \textbf{One-wayness:}	Given examples of $\sigma=\Sign(\skey,m)$ for attacker-chosen $m$ (but not given $\vkey$), an attacker cannot find $\vkey'$ for which $\Vrfy(\vkey', \sigma, m )$ returns true for any of the example $(\sigma,m)$.
	\item \textbf{Pseudorandomess:} The output of $\Sign(\skey,\cdot)$ looks random to an attacker that does not know $\vkey$ or $\skey$.
\end{itemize}

{\bf Privacy and Security Properties.} We discuss the privacy and integrity properties this protocol has in common with the earlier protocol, as well as some newer properties not achieved by the earlier protocol.\begin{itemize}
    \item {\em Privacy for negative users.} 
	 By the pseudorandomness property, the signatures broadcast by a user $u$ look pseudorandom. Beyond that, $u$ broadcasts two random strings and their view of the current time $t_i$ which is already known by any device hearing the broadcast.\footnote{We note that this information \emph{does} have the potential to be used for tracking if a user's device has some large systematic bias in its measurement of time.} Thus these broadcasts cannot be linked without knowledge of the internal state of $u$. As before, this privacy guarantee improves with the frequency of generating new IDs.
    \item {\em Privacy for positive users.} Upon reporting positive, the IDs broadcast by a user within a single day {\em can} be linked to each other. IDs broadcast on different days can be linked if the server does not hide which $\vkey$'s were reported together. Older IDs from days before the infection window and {\em newer} IDs from after the report cannot be linked with those in the infection window or with each other. Therefore, a positive user has the same guarantees as a negative user outside of the reported infection window.
    \item {\em Integrity guarantees.} It is infeasible for an attacker to upload to the server a value $\vkey$ which verifies an unreported ID that was broadcast by another user. This prevents the attacker from misreporting IDs of otherwise negative users and erroneously alerting their contacts.
    \item {\em Replay protection.} The incorporation of $t$ in each ID prevents an attacker from performing a replay attack where they gather IDs of legitimate users (to be tested positive) and re-broadcast the IDs at a later time to cause false beliefs of exposure. A $\vkey$ reported to the server cannot be used to broadcast further IDs that will be recognized by other users as matching that report.
    \item {\em Non-sensitive storage.} Because $H(r_i, t_i)$ looks random, the information intentionally stored by the app together with an ID does not reveal when the corresponding interaction occurred. (Of course, it may be possible to infer information about $t_i$ through close examination of how the ID was stored, e.g., where it was written in memory as compared to other IDs.)
\end{itemize}

\end{document}